\documentclass[journal]{IEEEtran}

\usepackage{tikz}
\newcommand*{\circled}[1]{\lower.7ex\hbox{\tikz\draw (0pt, 0pt)%
		circle (.5em) node {\makebox[1em][c]{\small #1}};}}

\usepackage{latexsym,bm,amsmath,amssymb}
\usepackage{cite}
\usepackage{threeparttable}
\usepackage{subfigure}
\usepackage{makecell}
\usepackage{multirow}
\usepackage{graphicx}
\usepackage{color}
\usepackage{enumerate}
\usepackage{algorithm}
\usepackage{algpseudocode}
\usepackage{pifont}
\usepackage{caption}
\usepackage{balance}
\usepackage{stmaryrd}

\def\ie{\textit{i.e.}}
\def\etc{\textit{etc. }}
\def\eg{\textit{e.g.}}
\def\etal{\textit{et al.}}

\def\OO{{\mathcal O}}

\pdfnoligatures
\emergencystretch=\maxdimen
\title{Challenges and Approaches for Mitigating Byzantine Attacks in Federated Learning}
\begin{document}

\author{
Junyu Shi$^{1}$,
Wei Wan$^{1,2,3,4}$,
Shengshan Hu$^{1,2,3,4}$,
Jianrong Lu$^{1,2,3,4}$,
Leo Yu Zhang$^5$

$^1$School of Cyber Science and Engineering, Huazhong University of Science and Technology\\
$^2$National Engineering Research Center for Big Data Technology and System\\
$^3$Services Computing Technology and System Lab\\
$^4$Hubei Engineering Research Center on Big Data Security\\
$^5$School of Information Technology, Deakin University\\

\{shijunyu220, wanwei\_0303, hushengshan, lujianrong\}@hust.edu.cn,\\
leo.zhang@deakin.edu.au
}
\maketitle

\begin{abstract}
Recently emerged federated learning (FL) is an attractive distributed learning framework  in which numerous wireless end-user devices can train a global model with the data remained autochthonous. Compared with the traditional machine learning framework that collects user data for centralized storage, which brings huge communication burden and concerns about data privacy, this approach can not only save the network bandwidth but also protect the data privacy. Despite the promising prospect, Byzantine attack, an intractable threat in conventional distributed network, is discovered to be rather efficacious against FL as well. In this paper, we conduct a comprehensive investigation of the state-of-the-art strategies for defending against Byzantine attacks in FL. We first provide a taxonomy for the existing defense solutions according to the techniques they used, followed by an across-the-board comparison and discussion. Then we propose a new Byzantine attack method called weight attack to defeat those defense schemes, and conduct experiments to demonstrate its threat. The results show that existing defense solutions, although abundant, are still far from fully protecting FL. Finally, we indicate possible countermeasures for weight attack, and highlight several challenges and future research directions for mitigating Byzantine attacks in FL.

\end{abstract}


\section{Introduction}
Ubiquitous intelligent devices equipped with advanced sensors (\eg, smartwatches, environmental monitoring devices) have brought us into the Internet of Things (IoTs) era, which connects the dispersive world into an interconnected system of intelligent networks. To make use of the data generated by these distributed devices, machine learning as a service (MLaaS)~\cite{mlaas} is becoming popular to assist users in refining their businesses. However, MLaaS usually needs to collect data from those devices and perform data analysis jobs in a centralized manner, which inevitably incurs two severe problems: high communication cost and privacy leakage~\cite{mlaasSecurity}. In IoTs, data is explosively generated every day, uploading all the raw data to a central server will bring a high burden to the bandwidth, especially in the wireless communication network. Besides, end-user devices usually contain a large amount of private information, such as location, identity, personal profiles, \etc Directly uploading local data to the server will raise great concerns on user privacy.

To address these issues, recently emerged federated learning (FL)~\cite{FL1,FL2,FL3} is a new computing paradigm that allows users to collaboratively compute a global machine learning model without revealing their local data. By distributing the model learning process to the end users (\eg, intelligent devices), FL constructs a global model from user-specific local models, ensuring that the users’ private data never leaves the devices. In this way, the bandwidth cost is significantly reduced and user privacy is well protected.

Despite the promising prospect, recent studies show that FL is highly susceptible to Byzantine attacks, where malicious users can falsify real models or gradients to damage the learning process, or directly poison the training data to make the global model learn the wrong information. Blanchard \etal~\cite{b1} have shown that just one baleful user can compromise  the convergence of the training and damage the performance of the ultimate global model. To address this issue, a mounting number of defense strategies against Byzantine attacks have been proposed to further safeguard FL~\cite{b2,b3,b4,b5,b6}. {\color{black}Although these research efforts have demonstrated their preliminary success in defeating Byzantine attacks, we emphasize that it is still far from practice to provide a full protection for FL. Protecting FL from Byzantine attacks that simultaneously considers the issues including efficiency, privacy, data distribution is an extremely challenging problem in the literature, especially when there still exist unknown attack surface in its standard process. Recently Mothukuri \etal~\cite{mothukuri2020survey} presented a comprehensive survey on the security and privacy of federated learning. However, it neither conducted experiments to evaluate existing schemes simultaneously, which is important to give a fair comparison, nor proposed any novel ideas to support its findings on the further work.}

In order to clearly demonstrate the vulnerability of existing Byzantine-robust FL schemes, this article first conducts a concise overview where an in-depth taxonomy is provided for the state-of-the-art defense strategies. We divide the existing defense solutions into four categories according to the principals they relied on for  anomaly detection, \ie, the distance based solution, the performance based solution, the statistics based solution, and the target optimization based solution. Then a comprehensive comparison is provided in terms of their advantages and disadvantages. After reviewing the literature, we
propose a new kind of Byzantine attack called weight attack to defeat those defense schemes. By making use of the flaw in the existing weight assignment strategy, our attack is much easier to put into practice while enjoying a high attack success rate. We further conduct experiments to validate the threat of our weight attack. Finally, we discuss the possible countermeasures, and highlight several stubborn challenges and future research directions for hardening the security of FL.

{\color{black}
In summary, we make the following contributions:
\begin{itemize}
  \item We provide a systematic review and comparison for state-of-the-art Byzantine-resilient federated learning schemes.
  \item We propose a new kind of Byzantine attack to show the feasibility of disabling existing defense methods, followed by experimental validations.
  \item We give an in-depth discussion for future work on enhancing the security of federated learning when facing Byzantine attacks.
\end{itemize}
}

\begin{figure*}
	\begin{center}
		\includegraphics[width= 1.8\columnwidth]{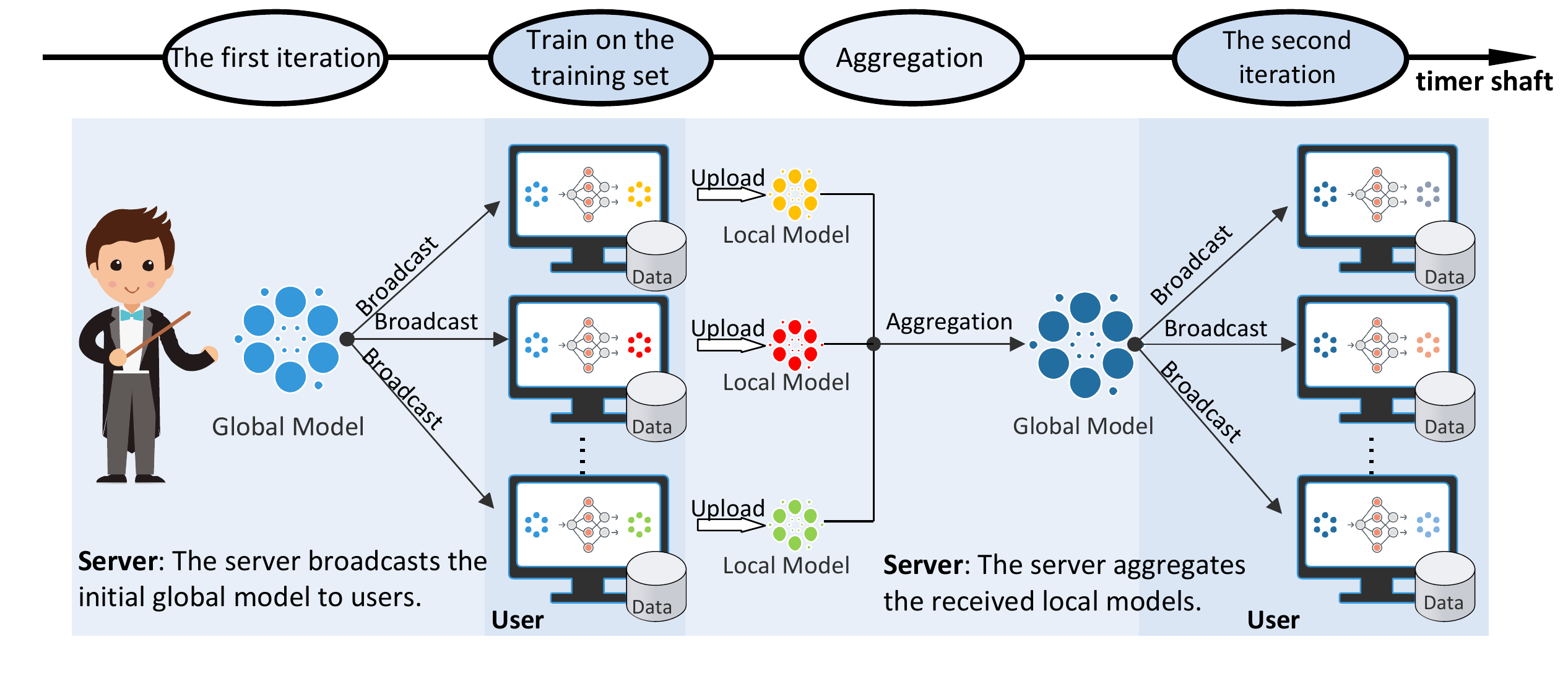}\\
		\caption{Architecture of federated learning}\label{fig:FL}
	\end{center}
\end{figure*}

\section{Preliminaries}

\subsection{Federated Learning}
{\color{black}In the conventional collaborative deep learning training framework, a powerful central server is usually required to gather users' training data. After receiving the data from users, the central server iteratively trains a deep neural network (DNN) model until it converges. In the end, users can download the DNN model and enjoy intelligent services. However, such training framework can easily lead to the leakage of user privacy, as users' private data is handed over to a third party.}

To address this issue, recently proposed federated learning (FL) is a distributed and privacy-protected architecture  in which  users collaboratively train and maintain a shared model under the architecture of a central server.  Fig.~\ref{fig:FL} depicts an overview of the standard FL architecture. In each iteration, the server first broadcasts a global model to a set of randomly chosen distributed users, each of which will then re-train a local model using their own data. After completing the local training, the users send the update (\ie, the local model) to the server for aggregation and generating a new global model. The iteration repeats until the global model converges.

However, FL still faces many technical challenges, such as  vulnerability to Byzantine attack, high communication overhead, dependence on the assumption of IID (\ie, independently and identically distributed data). In this paper, we mainly focus on defense schemes against Byzantine attacks.
%
%
%
%
\subsection{Byzantine Attack}
Recent works show that standard federated learning is vulnerable to Byzantine attacks carried out by faulty or malicious clients. Even if there is only one attacker, the model accuracy can drop from 100$\%$ to 0$\%$. For example, in an extreme case where an attacker knows the local updates of all benign clients, it only needs to set its update to the opposite of the linear combination of other normal updates to offset the effect of benign clients, then the accuracy of the aggregated global model can be reduced to 0$\%$ with a high possibility. We classify malicious attacks into two types based on which step in FL the malicious clients aim to breach:

\begin{enumerate}[1)]
\item Training data based attack: This kind of attack is also known as the data poisoning attack, which aims to mislead the global model by manipulating the local training data. In general, there are three main approaches for this attack:
	\begin{itemize}
	\item \textbf{Label flipping.} The attacker ``flips'' the  labels of its training data to arbitrary ones (\eg, via a permutation function).
	\item \textbf{Adding noise.} An attacker contaminates the dataset by adding noises to degrade the quality of models.
	\item \textbf{Backdoor trigger.} An attacker injects a trigger into a small area of the original dataset to cause the classifier misclassifying into the target category.
	\end{itemize}

\item Parameter based attack: This attack method involves altering local parameters (\ie, gradient or model) so that the central server aggregates a corrupted global model. There are two ways to modify the parameters:
	\begin{itemize}
	\item \textbf{Nonrandom modification.} Modifying the direction or the size of the parameters based on the model faithfully learned from the local dataset, \eg, flipping the signs of local iterates and gradients, or enlarging the magnitudes.
	\item\textbf{Random modification.} Modifying the parameters directly, \eg, randomly sampling a number from the Gaussian distribution and treating it as one of the parameters of the local model.
	\end{itemize}

\end{enumerate}

\section{Existing defense schemes}
Depending on the principles that the server relies on for detecting or evading anomalous updates, the existing defense schemes can be divided into four categories: the distance based defense schemes, the performance based defense schemes, the statistics based defense schemes and the target optimization based defense schemes.
\subsection{The Distance based Defense Schemes}
This kind of defense schemes aims to detect and discard bad or malicious updates by comparing the distances between the updates. The update which is apparently far away from the others is regarded as  malicious. These schemes are usually easy to implement .

 Blanchard \etal~\cite{b1} proposed Krum and its variant, called Multi-Krum. In Krum, the central server chooses only one update that is closest to its neighbors to update the global model, and discards all the other updates, whereas Multi-Krum chooses multiple updates and computes the mean to update the global model. Similar to Multi-Krum, FABA \cite{b2} aims to remove the outliers in the uploaded gradients  by discarding the gradients that are far away from the mean gradient. However, both Multi-Krum and FABA need to know the number of malicious clients in advance, which makes them difficult to be applied to practical applications. To get rid of this limitation, FoolsGold \cite{b9} uses cosine similarity to identify malicious updates and assigns them a low weight to reduce their impact on the global model. In their viewpoint, the updates from attackers have nearly the same direction, thus the cosine similarities between abnormal updates should be extraordinarily large. Based on this observation, the central server can find an abnormal update and assigns them low weights. Cao \etal~\cite{b13} proposed Sniper, which utilizes Euclidean distances between local models to construct a graph, based on which a set of updates will be selected for aggregation. The above defenses apply to the scenario where malicious updates are far from each other (\ie, non-sybil case), however, attackers may collude to upload extremely similar or even identical updates to make the attack more covert (\ie, sybil case). In light of this, Wan \etal~\cite{WeightAttack} proposed a density-based anomaly algorithm to detect similar updates. Nevertheless, the defense only focuses on the sybil case and ignores the non-sybil case. To deal with both sybil and non-sybil cases simultaneously, Wan \etal~\cite{MAB-RFL} proposed a two-pronged defense called MAB-RFL, which first discards the updates that are excessively similar in direction through graph theory, aiming to cope with the collusion attack. Then it utilizes principal components analysis (PCA) to extract the key parameters lie in the updates, because benign and malicious updates are easier to distinguish (\eg, through agglomerative clustering) in the new low-dimensional parameter space. In this way, the non-sybil challenge can be settled.
 
\subsection{The Performance based Defense Schemes}
In this category, each update will be evaluated over a clean dataset provided by the server, such that any update that performs poorly will be assigned low weights or removed directly.

Li \etal~\cite{b3} leveraged a pre-trained autoencoder to evaluate the performance. For a benign model update, the autoencoder will output a vector that is similar to the input, but an abnormal update will generate a large gap. However, training an autoencoder is time-consuming, and it is difficult to get the training set that includes sufficient benign model updates. In contrast, Zeno \cite{b4} only requires a small validation set on the server-side. Specifically, Zeno computes a score for each candidate gradient with the validation set. The score is composed of two parts: the estimated descendant of the loss function, and the magnitude of the update. A higher score of the update implies  better performance, indicating a higher probability of being reliable. Nevertheless, Zeno requires the knowledge about the number of attackers. To address this problem, Cao \etal~\cite{b14} proposed a Byzantine-robust distributed gradient algorithm, which can filter out information received from the compromised clients by computing noisy gradient with a small clean dataset. The update whose distance with the noisy gradient satisfies a pre-defined condition will be accepted. However, the defense proposed by Cao \etal~\cite{b14} heavily relies on the setting of hyper-parameters, which may difficult to find when putting it into practice. In light of this, Cao \etal~\cite{FLTrust} proposed FLTrust, which utilize ReLU-clipped cosine similarity between each local update and the server update (calculated based on the clean dataset collected by the central server) to allocate trust score for the local update. A reliable update will obtain a high trust score due to its consistency with the golden standard (\ie, the server update), while a malicious update will obtain a 0 trust score. The higher the trust score, the greater the weight of the corresponding local update 
when aggregation.

\subsection{The Statistics based Defense Schemes}
Schemes in this category exploit the statistical characteristics of uploaded updates, such as the median or mean, to circumvent abnormal updates to get a robust one.

Yin \etal~\cite{b7} proposed two robust distributed gradient descent algorithms by computing coordinate-wise median and the coordinate-wise trimmed mean of all local updates in each dimension, respectively. Meanwhile, Xie \etal~\cite{b8} proposed three aggregation rules: geometric median, marginal median, and ``mean around median''. The geometric median intends to find a new update that minimizes the summation of the distances between the update and each local update. The marginal median is similar to the coordinate-wise median proposed in \cite{b7}. The ``mean around median'' takes the average of the values near to the median for each dimension of local update to obtain a new global update. However, the scheme in \cite{b8} needs to call a secure average oracle many times, which incurs expensive computational overhead. In light of this, Pillutla \etal~\cite{b10} proposed RFA (Robust Aggregation for Federated Learning) by computing the geometric median with a alternating minimization approach, which calls the secure average oracle for only three times. Bulyan \cite{b11}, which is modified based on \cite{b7}, executes a robust detection algorithm, such as Multi-Krum, before the aggregation with trimmed mean. The experimental results show that Bulyan performs better than using Krum alone. Mu{\~{n}}oz{-}Gonz{\'{a}}lez \etal~\cite{b6} proposed AFA (Adaptive Federated Averaging), which separately computes the cosine similarity between each local model and the global model, and discards bad models based on the statistical distribution of the median and the average of these cosine similarities. Xie \etal~\cite{b12} also used  trimmed mean as the aggregation rule, and they further proposed a moving-average method, which considers global models in two successive rounds.
\subsection{The Target Optimization based Defense Schemes}
The target optimization based defense schemes refer to optimizing a different objective function to improve the robustness of the global model.

Li \etal~\cite{b5} proposed RSA (Byzantine-Robust Stochastic Aggregation), which adds a regularization term to the objective loss function, such that each regular local model is forced to be close to the global model. As far as we know, this is the only work in this category so far.

\begin{table*}[!t]
	\centering
	\newcolumntype{M}[1]{>{\centering\arraybackslash}m{#1}}
	\renewcommand{\arraystretch}{1.2}
	\addtolength{\tabcolsep}{-4pt}

	\begin{tabular}{ |M{8em}||M{7em}|M{7em}|M{10em}|M{10em}|M{6em}|M{10em}| }
		\hline
		\textbf{Solution} & \textbf{Category} & \textbf{Target attack} & \textbf{The number of attackers} & \textbf{Model accuracy} & \textbf{Data distribution} & \textbf{Time complexity} \\
		\hline
		Multi-Krum \cite{b1} & Distance based & Data/parameter based & Less than 50$\%$ & Medium & IID & $\OO(K^{2}d)$ \\
		\hline
		FABA \cite{b2} & Distance based & Data/parameter based & Less than 50$\%$ & Medium & IID & $\OO(K^{2}d)$ \\
		\hline
		Sniper \cite{b13} & Distance based & Data based & Less than 50$\%$ & Medium & IID & $\OO(K^{2}d)$ \\
		\hline
		FoolsGold \cite{b9} & Distance based & Data/parameter based & No limitation & Medium (FoolsGold) High (FoolsGold+Multi-Krum) & IID/non-IID & $\OO(K^{2}d)$ \\
		\hline
		Wan \etal \cite{WeightAttack} & Distance based & Parameter based & Less than $50\%$ & High & IID & $\OO(K^{2}d)$ \\
		\hline
		MAB-RFL \cite{MAB-RFL} & Distance based & Data/parameter based & Less than $50\%$ & High & IID/non-IID & $\OO(K^{2}d)$ \\
		\hline
		Li \etal \cite{b3} & Performance based & Data/parameter based & Less than 50$\%$ & High & IID/non-IID & $\OO(Kd)$ \\
		\hline
		Zeno \cite{b4} & Performance based & Data/parameter based & No limitation & High & IID/non-IID & $\OO(Kd)$ \\
		\hline
		Cao \etal \cite{b14} & Performance based & Data/parameter based & No limitation & High & IID & $\OO(Kd)$ \\
		\hline
		FLTrust \cite{FLTrust} & Performance based & Data/parameter based & No limitation & High & IID/non-IID & $\OO(Kd)$ \\
		\hline
		AFA \cite{b6} & Statistic based & Data/parameter based & Less than 50$\%$ & High & IID & $\OO(K^{2}d)$ \\
		\hline
		GeoMed \cite{b8} \newline MarMed \cite{b7,b8} \newline Trimmed mean \cite{b7,b8} & Statistic based & Data/parameter based & Less than 50$\%$ & High & IID/non-IID & $\OO(KdlogK)$ \\
		\hline
		Bulyan \cite{b11} & Statistic based & Data/parameter based & Less than 50$\%$ & Medium & IID  & $\OO(K^{2}d)$ \\
		\hline
		SLSGD \cite{b12} & Statistic based & Data/parameter based & Less than 50$\%$ & Medium & IID/non-IID  & $\OO(KdlogK)$ \\
		\hline
		RFA \cite{b10} & Statistic based & Data/parameter based & Less than 50$\%$ & Medium & IID  & $\OO(Kd)$ \\
		\hline
		RSA \cite{b5} & Target optimization based & Data based & No limitation & High & IID/non-IID & $\OO(Kd)$ \\
		\hline
	\end{tabular}
	
	\caption{A comprehensive comparison of Byzantine-robust FL method. The model accuracy represents the prediction accuracy of the defense scheme, and ``Medium'' and ``High'' indicate that the accuracy is below and close to that of non-attacker case, respectively. ``IID'' means that the users’ local data sets are independently and identically distributed, while ``non-IID'' indicates otherwise. $K$ denotes the number of users and $d$ denotes the model size.}
	\label{tab:compALL}
\end{table*}

\section{A comprehensive comparison}
Based on the above discussion, we know that the distance based defense schemes usually rely on the assumption that the parameter distribution of a malicious attacker is scattered and deviates from the benign ones, so it is only suitable to resist the attacks that generate evident abnormal parameters. For example, the label flipping attack can easily cause significant changes in parameters. However, it is obvious that such defense schemes perform poorly when the attack causes faint changes.

The performance based defense schemes detect anomalous updates by directly verifying their performance, which is much more reliable than other solutions. For example, Zeno \cite{b4} is superior to Krum under both the data based attack (\eg, sign flipping) and the parameter based attack (\eg, random gradient). However, it relies on a clean auxiliary data set for examination, which hampers its practicability.  Besides, scheme \cite{b3} has a high time complexity since the time-consuming pre-training is required when using the auto-encoder.

The statistics based defense schemes rely on computing median or mean to evade abnormal parameters, making it only suitable to the situation where the number of malicious users is less than half of the total users, otherwise legitimate updates will be left out when  malicious  updates dominate. In comparison, the target optimization based defense scheme enjoys a high efficiency. For instance, according to the experimental results in \cite{b5}  conducted on the MNIST dataset, the time cost of RSA is around 45s, while Median, Krum, and GeoMed cost about 50s, 62s, and 127s, respectively.

TABLE~\ref{tab:compALL} presents a comprehensive comparison among existing solutions.
We can see that various defense strategies have already been proposed, each of which has its own merits and demerits, and many tough issues have been discussed in detail, such as non-IID condition, more than 50\% attackers, \etc Nevertheless, we emphasize that it is still far from practice to deploy a secure framework for FL in the presence of Byzantine attacks.
 {\color{black}Fully protecting FL that integrally considers the issues including efficiency, data privacy, and data distribution is an extremely challenging problem, especially when there still exist many attack surfaces in its standard process. In the next section, we will present a newly found attack approach to support our observation. }
%
%

%

\section{Weight attack}
In this section, we propose a new attack approach called weight attack to circumvent those defense schemes. The key idea lies in manipulating the drawbacks of the weight assignment strategy that have not received enough attention yet. {\color{black}Our attack is simple and easy to carry out in practice, and performs well even when those defense schemes were deployed.}

\subsection{System Overview}\label{Sec:SysOverview}
In the standard federated learning setting, when aggregating updates in each iteration, the weight assigned to each update totally depends on the size of the local training data set~\cite{b1,b2,b4,b12,b13,b14}. The central server has no authority or  effective means to check the sizes and quality of clients' training data, due to the privacy reasons. Therefore, the local data set size is declared by the clients themselves without any verification.

Based on this observation, any malicious client can arbitrarily lie about its data set size for gaining a high weight.
According to the way the attackers declare their training data set sizes, we consider the following two simple misreport cases:
\begin{enumerate}[1)]
	\item The attackers' training data set sizes are much smaller than that of the regular clients, but they declare that they have similar sizes with the regular clients.
	\item The attackers and the regular clients have similar training data set sizes, but the attackers declare that their training set sizes are much larger than that of the regular clients.
\end{enumerate}

Obviously, it is very easy for attackers to initiate such attack since the server cannot examine the clients' declarations. Next, we briefly introduce the specific process of the weight attack.

\subsection{Algorithm Design}
\textbf{Step 1.} The central server broadcasts the global model to each selected client.

\textbf{Step 2.} Each client, including the attackers, re-trains the global model based on its local training data set.

\textbf{Step 3.} The clients send the updates to the server. And the attackers misreport their data set sizes, while the regular clients faithfully report their training set sizes.

\textbf{Step 4.} The central server aggregates the received updated models to obtain a new global model and repeats steps 1 to 4 until the global model converges.

\begin{figure*}[t]
	\centering
		\subfigure[]{
		\includegraphics[width=0.45\columnwidth]{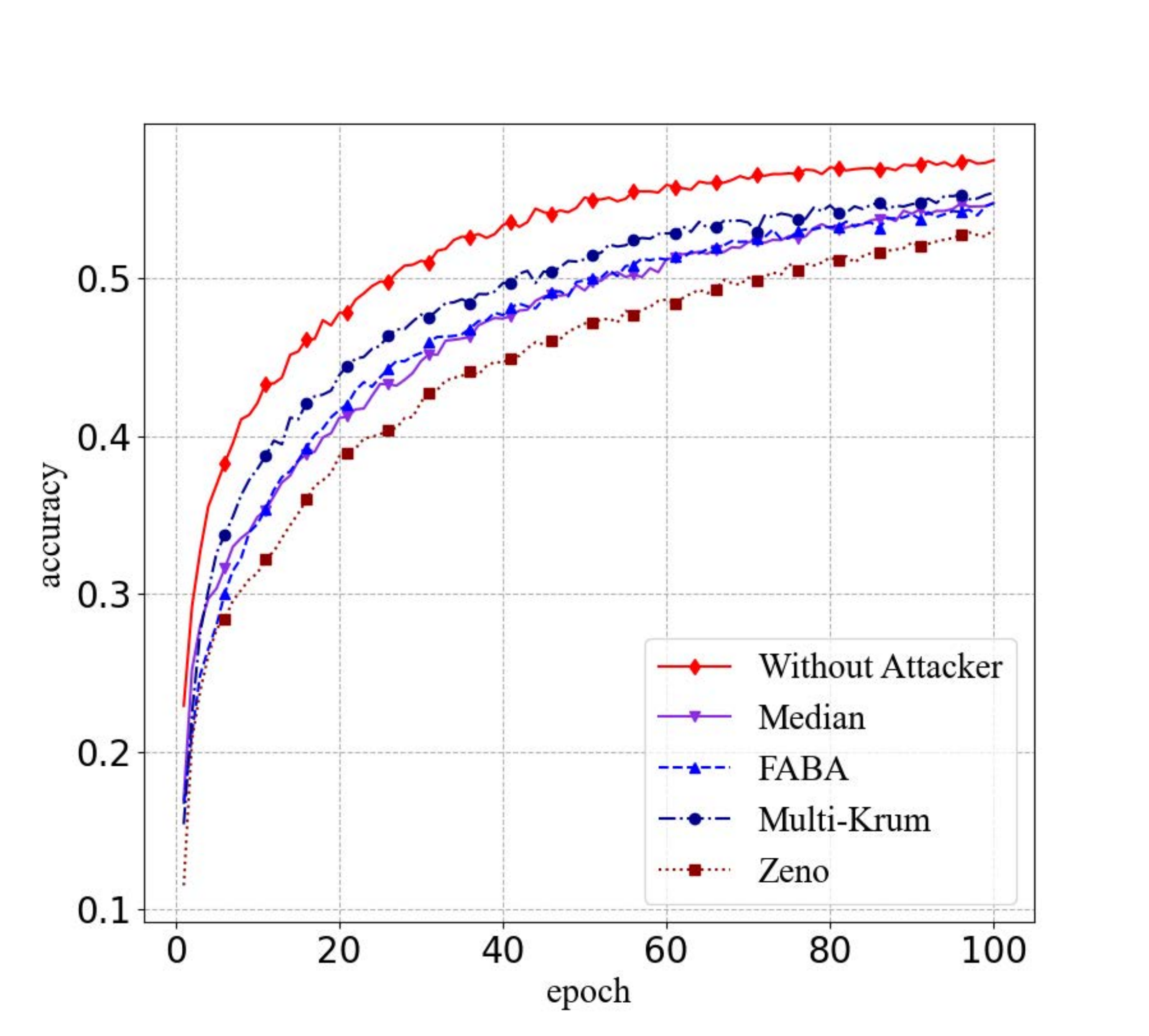}}
	\subfigure[]{
		\includegraphics[width=0.45\columnwidth]{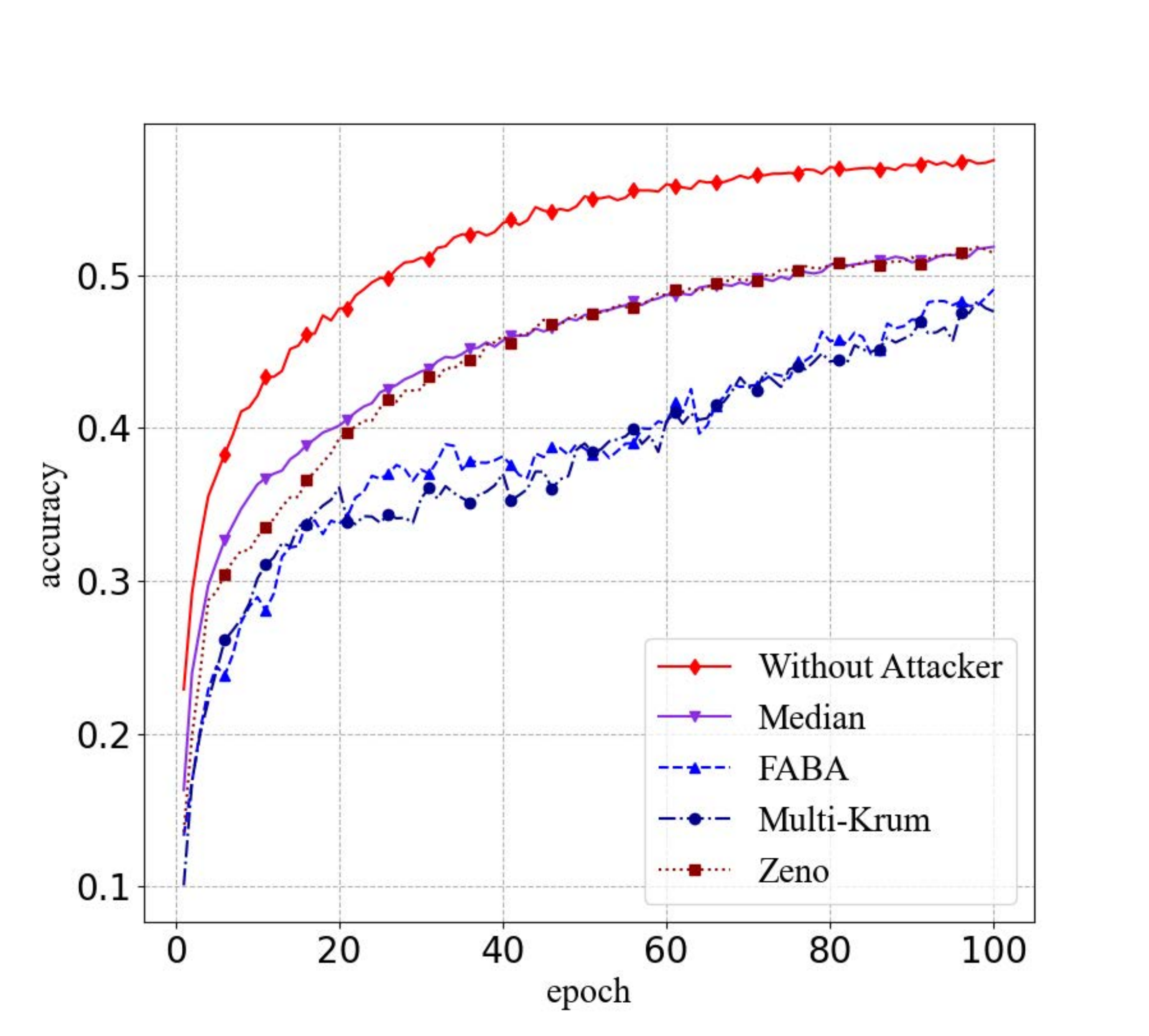}}
	\subfigure[]{
		\includegraphics[width=0.45\columnwidth]{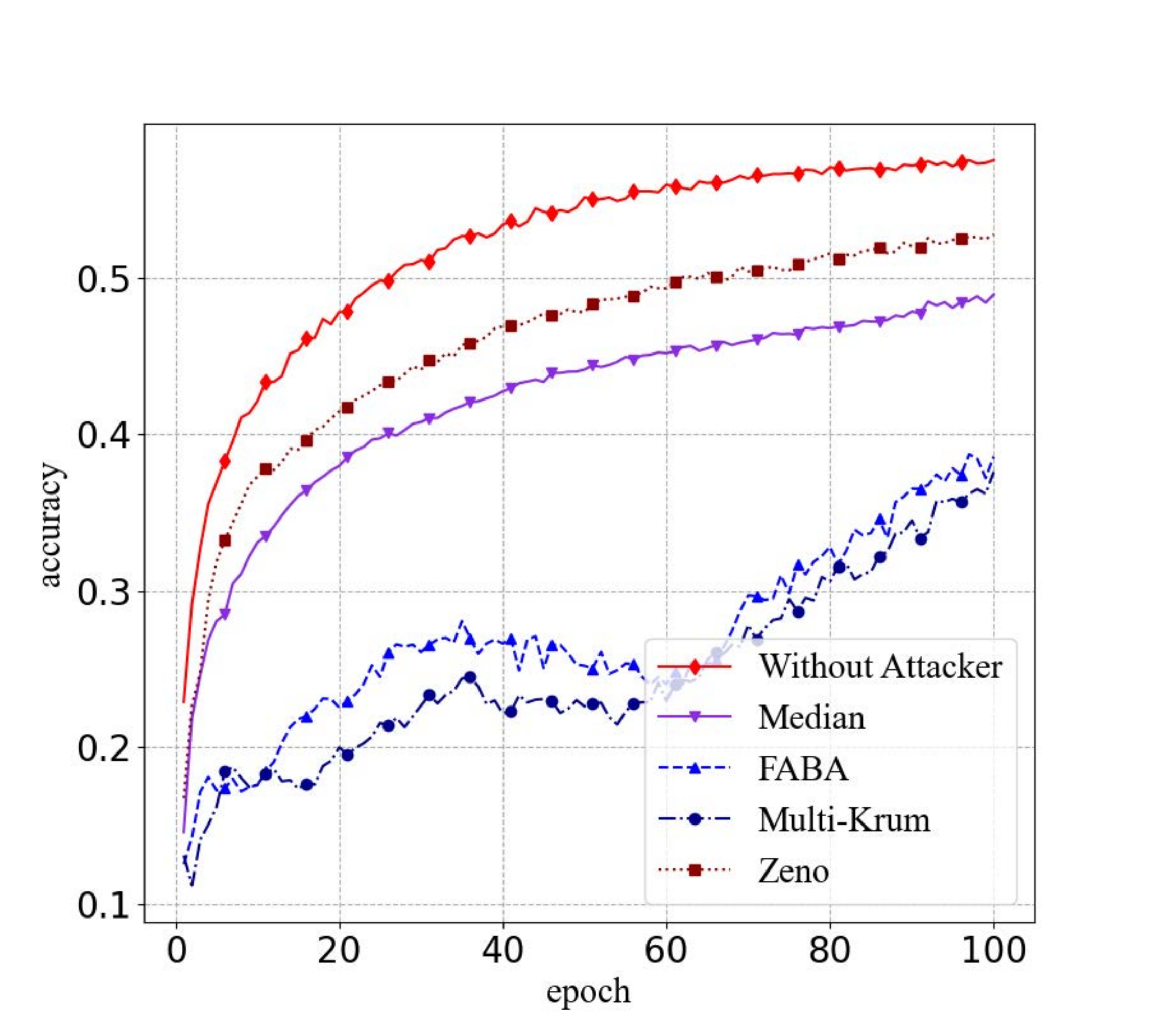}}
	\subfigure[]{
		\includegraphics[width=0.45\columnwidth]{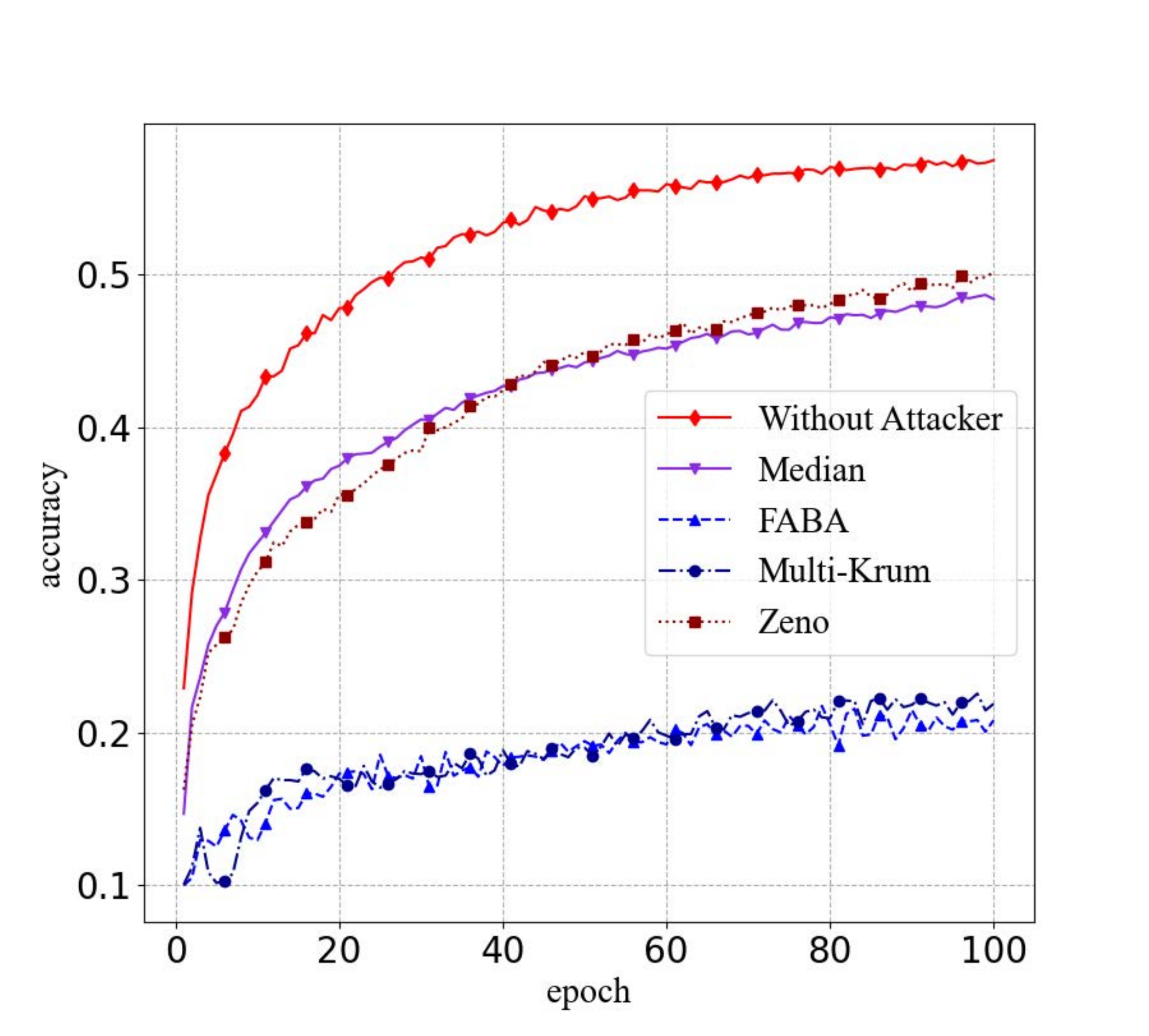}}
	\caption{The accuracy of existing defense schemes under weight attack with: (a) 20$\%$ attackers, (b) 30$\%$ attackers, (c) 40$\%$ attackers, (d) 50$\%$ attackers}
	\label{fig:accuracy}
\end{figure*}

\section{Experiments and evaluation}
In this section, we conduct experiments to show the effectiveness of the weight attack. The experiments are implemented with Tensorflow on CIFAR-10 image classification dataset, which is composed of 50K images for training and 10K images for testing. We use CNN with 2 convolutional layers followed by 2 fully connected layers. Since our purpose is to examine the effectiveness of the weight attack, we omit the client selection process  and assume that there are  20 clients in total and all of them are selected in each round.

We only consider the first misreport case defined in Section~\ref{Sec:SysOverview} since it is more practical than the second case. This is because if a bad node claims to have a larger dataset size than that of regular clients, the central server is easier to find abnormality and thus require the node re-upload the local update or prove that the update is indeed derived from the claimed dataset size.
In the first misreport case, however, all the dataset sizes are similar, the central server cannot tell which nodes might be malicious just from the dataset sizes.

{\color{black}We test four typical defense schemes (\ie, Multi-Krum, FABA, Zeno, and Median), and evaluate the weight attack with 4, 6, 8, and 10 attackers among 20 clients, \ie, there are 20$\%$, 30$\%$, $40$\% and 50$\%$ attackers, respectively. We set the training data set size (\ie, number of images) to be 2500 and 100 for regular clients and attackers, respectively. But they gain equal weight for each update on the server side. And it should be noted that the clients' data is allocated in an IID way. Specifically, we randomly scramble all the data and allocate the appropriate amount of data based on the dataset size of each client. As a comparison, we also consider the case without attackers.}

In Fig.~\ref{fig:accuracy}, we observe that with 20$\%$ attackers, Multi-Krum, FABA, and Median have similar performance, and their accuracy is lower than the case without attacker, while Zeno performs slightly worse than the other three schemes. In the case of 30$\%$ attackers, the accuracy of Multi-Krum and FABA is significantly affected. In comparison, Median and Zeno performs better than Multi-Krum and FABA, and their accuracy is about 52$\%$. When there are 40$\%$ attackers, Zeno performs similar to the case of 30$\%$ attackers, while the accuracy of the other three schemes is further reduced. In the case of 50$\%$ attackers, Multi-Krum and FABA cannot converge, and both of them have low accuracy (\ie, 20$\%$). Although Zeno and Median perform better, their accuracy is still 10$\%$ lower than the case without attacker.

\begin{figure}[t]
	\centering
	\subfigure[]{
		\includegraphics[width=0.45\columnwidth]{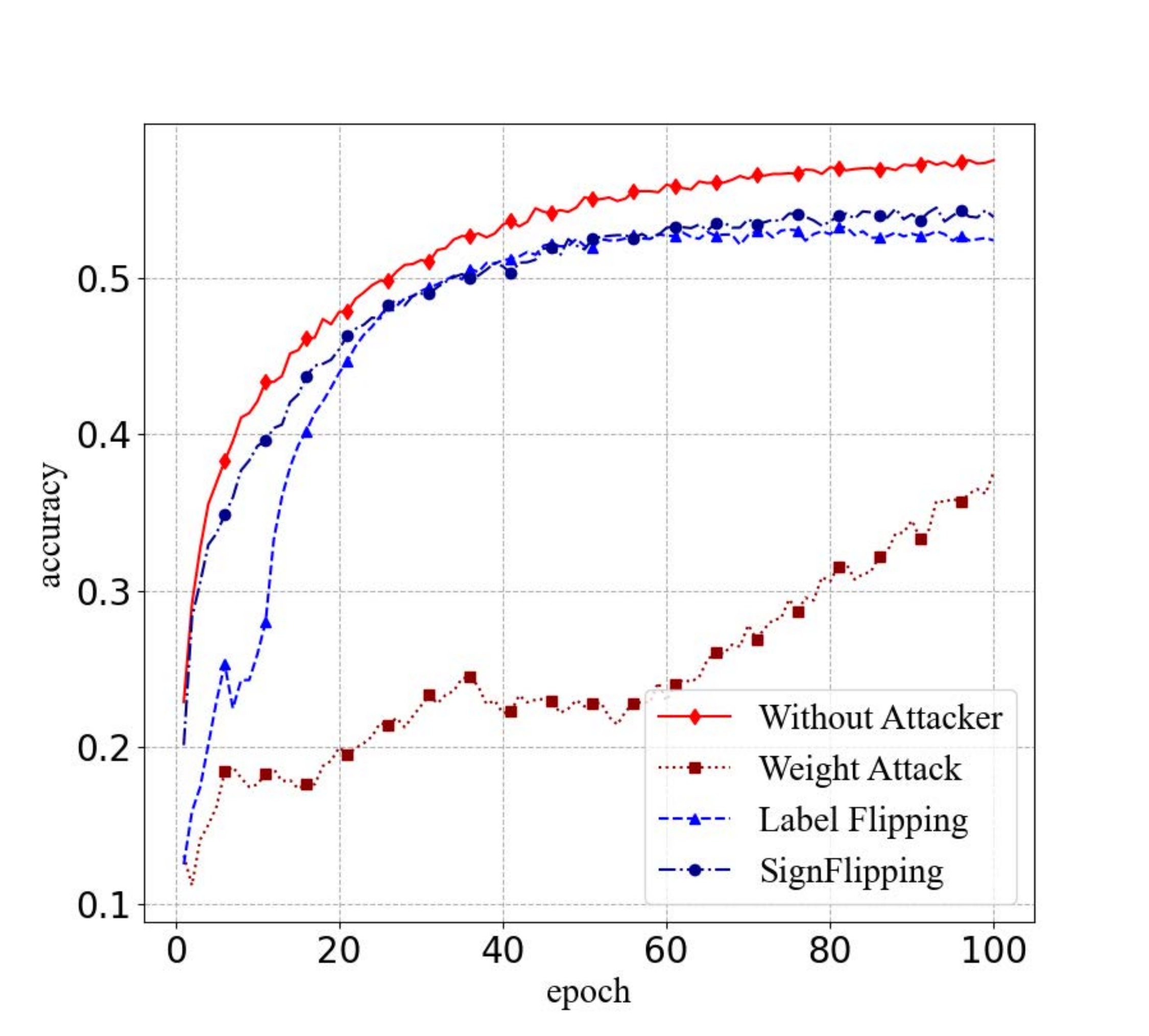}}
	\subfigure[]{
		\includegraphics[width=0.45\columnwidth]{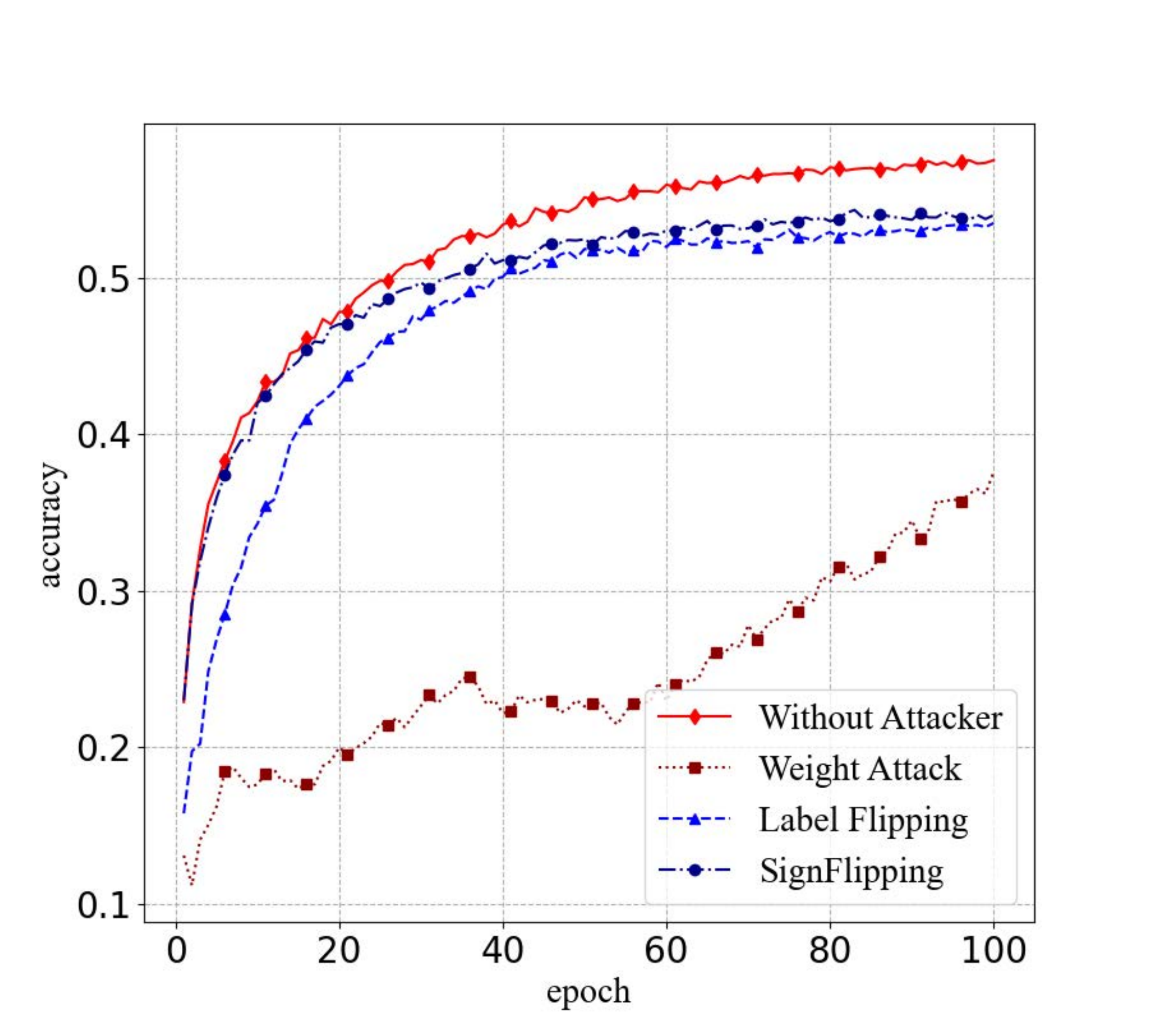}}
	\caption{The comparison between weight attack and existing Byzantine attacks under (a) Multi-Krum, (b) FABA}
	\label{fig:attackscompare}
\end{figure}

We also give a comparison between our weight attack with two typical Byzantine attacks: label flipping attack and sign flipping attack.
In the sign flipping attack, after obtaining the local model, the attacker multiplies it with  a negative number. We set the negative number to be $-4$ in our experiments, which is also adopted in existing works. The experiments are conducted on CIFAR-10 data set, and 40$\%$ participants are malicious among 20 clients. Note that Multi-Krum and FABA are used as the defense.
As shown in Fig.~\ref{fig:attackscompare}, label flipping attack and sign flipping attack have little effect on  the accuracy when Multi-Krum or FABA is deployed, and both of the attacks reduce the accuracy by 2$\%$-5$\%$. On the contrary, the weight attack has a high attack success rate, which can reduce the accuracy by 20$\%$.

From the above experiments, we can conclude that the weight attack can indeed decay the existing defense schemes, especially in reducing the prediction accuracy of the global model or even preventing it from converging.

\section{Possible solutions to weight attack}
Existing defense solutions are not able to mitigate the weight attack. The main difficult lies in the fact that the server cannot directly examine the quality of the clients' local data sets.  Next, we discuss some possible countermeasures.

Although the distance based schemes such as Multi-Krum and FABA fail to resist the weight attack, we think it is still a promising solution. The reason why both Multi-Krum and FABA fail is that they are inclined to exclude updates that are far from the overall distribution. We hold the viewpoint that by analyzing the distribution of local updates, designing a new distance based strategy is able to directly evade the ``bad'' updates.

Besides, as shown in our experiments, the performance based defense scheme, such as Zeno, performs much better than other schemes. We think this kind of defense scheme can do better in the future, because the most straightforward way to determine whether an update is benign or malicious is to examine its performance. The ``bad'' updates generated from the weight attack are sure to act differently  when the clean test data set is well designed for some experiments.

As for the statistics based and the target optimization based defense schemes, we firmly believe that by fully exploiting the statistical characteristics of local updates or selecting a good loss to optimize the objective function, they can mitigate the weight attack effectively as well.

\section{Challenges and research directions}

Although the Byzantine attacks in FL have attracted much research interest and great efforts have been devoted to designing a secure FL scheme, there are many open problems that need to be further investigated. In this section, we outline some challenges and research directions which we believe are of great significance for defending against Byzantine attacks.

\subsection{Fair Reward Distribution in FL}

The success of the weight attack relies on the assumption that the server assigns the weight to updates according to the local training data set size of each client. It enables attackers to arbitrarily claim their workload in the local training by falsely reporting their data set size. Moreover, even if the server has deployed effective detection methods, if any, to discard bad updates generated from the weight attack, the clients may also tend to misbehave (\eg, being lazy) since they will finally share the same global model, no matter how many  data sets are used for local training or how much computing power are devoted.

In light of this, we emphasize that it is of significant importance to design a fair incentive mechanism before putting FL into practice. Apart from the training data set size, more metrics (\eg, model quality, computing resources, past behavior) should be included to evaluate the contribution of each client such that they will be fairly rewarded. As a result, each client is incentivized to do correct computations. This can not only discourage the weight attack, but also motivate clients to contribute more resources to speed-up the local training.

\subsection{Defending Attacks with Privacy Protection}
The primary goal of federated learning is to protect users' privacy by requiring them to upload local updates instead of their training data. However, recent research~\cite{ModelInversion} has shown that deep models will also reveal private information about the training data. Existing defense schemes mainly focus on mitigating Byzantine attacks, but ignore the possible privacy leakage through the updates.

Hence, it is necessary to consider privacy protection while defending against Byzantine attackers simultaneously. The most straightforward solution is letting the client encrypt local updates before uploading them to the central server, who can carry out a defense scheme to detect anomaly over encrypted data. Then the central server broadcasts the encrypted global model to clients for decryption and proceeds to the next iteration. The main challenge for this method lies in designing a secure computation protocol that can effectively detect anomaly while protecting the privacy of update, without affecting the performance of the final global model at the same time. Cryptographic tools such as homomorphic encryption or garbled circuit can provide accurate computations over encrypted data, but they will bring a high computation overhead. Other privacy-enhancing techniques like differential privacy, or hardware-based trusted execution environment enjoy a high efficiency, but they may cause loss on the model accuracy or cannot fully protect the private information. A trade-off between efficiency, security, and privacy needs to be carefully considered for specific application scenarios.

\subsection{Shielding FL in Label Deficiency Scenario}
Nearly all the existing Byzantine-robust defenses focus on the label sufficiency scenario, where each client is equipped with fully labeled data. In other words, they concentrate on supervised learning tasks. However, in real-world scenarios, clients' data may be slightly labeled or even totally unlabeled due to the lack of expertise and motivation of the clients to label their own data. Although a few works~\cite{FSSL1,FSSL2,FSSL3} begin to research how to train a high-quality global model in the label deficiency scenario, the security problem is overlooked.

One may expect to directly apply existing defenses to the label deficiency FL scenario, however, this may not work. As we know, one key reason why detecting malicious updates in non-IID scenario is difficult lies in the tremendous differences between benign updates, which makes it intractable to distinguish between benign updates and malicious ones. Worse still, the differences will be amplified in the label deficiency scenario. Because the pseudo-labels generated by the local model may be incorrect, especially in the early iterations, which may be caused by the low-quality of the received global model. On the other hand, one usually sets a confidence threshold to make the pseudo-labels more reliable, so only those highly confident data will be used for training, making the local updates biased. 

Therefore, shielding FL in label deficiency scenario is challenging and significant. We argue that variance reduction techniques (\eg, resampling~\cite{resampling}, grouping~\cite{groupingFSSL} and momentum~\cite{momentum}) will be promising to solve the problem, as they have shown splendid performance in coping with the data heterogeneity challenge. By reducing the differences between benign updates, an IID-like distribution can be built, which is more conducive to detecting malicious updates for the server.

%
%
%

\section{Conclusion}
The advance of federated learning has given researchers a new direction in addressing the security and privacy issues of distributed training. Mitigating Byzantine attacks is important for securing federated learning. In this article, we review existing solutions for defending against Byzantine attacks. After a comprehensive comparison and discussion, we propose a new attack method that can pose threats to existing defense schemes, supported by our experimental results. Finally, we indicate several challenges and the future research direction of FL.

\section*{Acknowledgments}
Shengshan’s work is supported in part by the National Natural Science Foundation of China (Grant Nos. 62002126, U20A20177), and Fundamental Research Funds for the Central Universities (Grant No. 2020kfyXJJS075). Leo’s work is supported in part by the National Natural Science Foundation of China (Grant No. 61702221). Wei's work is supported in part by Ant Group. Shengshan is the corresponding author.
\bibliographystyle{IEEEtran}
\bibliography{reference}

\end{document}